# Plasmon-enhanced two photon excited emission from edges of one-dimensional plasmonic hotspots with continuous-wave laser excitation

Tamitake Itoh[1]*, Yuko S. Yamamoto[2]

[1]Health and Medical Research Institute, National Institute of Advanced Industrial Science and Technology (AIST), Takamatsu, Kagawa 761-0395, Japan

[2]School of Materials Science, Japan Advanced Institute of Science and Technology (JAIST), Nomi, Ishikawa 923-1292, Japan

*Corresponding author: tamitake-itou@aist.go.jp

**ABSTRACT**

One-dimensional junctions between parallel and closely arranged multiple silver nanowires (NWs) exhibit a large electromagnetic (EM) enhancement factor ($F_R$) owing to both localized and surface plasmon resonances. Such junctions are referred to as one-dimensional (1D) hotspots (HSs). This study found that two-photon excited emissions, such as hyper-Rayleigh, hyper-Raman, and two-photon fluorescence of dye molecules, are generated at the edge of 1D HSs of NW dimers with continuous-wave near-infrared (NIR) laser excitation and propagated through the 1D HSs; however, they were not generated from the centers of 1D HSs. Numerical EM calculations showed that $F_R$ of the NIR region for the edges of 1D HSs was larger than that for the centers by approximately $10^4$ times, resulting in the observation of two-photon excited emissions only from the edge of 1D HSs. The analysis of the NW dimer gap distance dependence of $F_R$ revealed that the lowest surface plasmon (SP) mode, compressed and localized at the edges of the 1D HSs, was the origin of the large $F_R$ in the NIR region. The propagation of two-photon-excited emissions was supported by higher-order coupled SP mode.

**I. Introduction**

The spontaneous emission rates of coupled systems between localized plasmon (LP) polaritons and molecular excitons are enhanced by the Purcell effect. This is expressed as $F = \dfrac{Q(\lambda/n)^3}{4\pi^2 V_P}$; where, $Q$ is the quality factor of the LP resonance, $\lambda$ and $n$ are the light wavelength and refractive index around the system, respectively, and $V_P$ is the mode volume of the coupled system [1]. Owing to the extremely small $V_P$ of a nanogap between metallic nanoparticle (NP) dimers, a molecule located inside the nanogap is subjected to a large electromagnetic (EM) enhancement factor $F_R$, which is expressed by the radiative portion of $F$ as

$$F_R = \frac{F \Delta\omega_R}{\Delta\omega_R + \Delta\omega_{NR}} \left( = \left|\frac{E_{loc}}{E_{in}}\right|^2 \right), \quad (1)$$

where $\Delta\omega_R$ and $\Delta\omega_{NR}$ are the radiative and nonradiative decay rates of the LP resonance, respectively, and $E_{in}$ and $E_{loc}$ are the amplitudes of the incident and local electric fields inside the nanogap, respectively [1]. The maximum $F_R$ reaches $<10^5$ at the nanogaps [2,3]; such nanogaps are referred to as hotspots (HSs). HSs have received considerable attention because they exhibit various interesting phenomena, such as single-molecule surface-enhanced Raman scattering (SERS) under resonant condition [4-7], ultrafast surface-enhanced fluorescence (ultrafast SEF) [8,9], vibrational pumping [10], field gradient effect [11], and strong coupling between plasmons and molecular excitons [12-15]. However, the value of $V_P$ of HS is only several tens of nm$^3$ [16], which induces serious instability [17]. To resolve this problem while maintaining a large $F_R$, several HSs have been developed both theoretically and experimentally [18]. Silver NWs are among the most promising plasmonic materials for HSs because of the synergistic effects between LP and surface plasmon (SP) resonances [19]. We developed one-dimensional HSs (1D HSs) wherein $V_P$ was extended by approximately $10^4$ times using the gaps or crevasses between silver NW dimers [20]. We also revealed that the $F_R$ of 1D HSs was generated by both superradiant and subradiant LP resonances [21], and that SPs supported the propagation of SERS and SEF light along a 1D HS [22].

One of the most promising applications of 1D HS is nonlinear optical spectroscopy, which typically requires high $E_{in}$ fields of ultrafast laser pulses owing to the small nonlinear molecular susceptibilities [23,24]. The excited polarization $P_{in}$ is expressed as a power series in $E_{in}$ of Eq. (1) as

$$P_{in} = \varepsilon_0 \left[ \chi^{(1)} E_{in} + \chi^{(2)} E_{in}^2 + \chi^{(3)} E_{in}^3 + \cdots \right] = P_{in}^{(1)} + P_{in}^{(2)} + P_{in}^{(3)} + \cdots \quad (2),$$

where $\chi^{(2)}$ and $\chi^{(3)}$ with $P_{in}^{(2)}$ and $P_{in}^{(3)}$ are the second- and third-order nonlinear susceptibilities with excited polarizations, respectively [25]. By assuming $\chi^{(1)}$ as unity, the typical values of $\chi^{(1)}$, $\chi^{(2)}$ and $\chi^{(3)}$ are 1, 2 × 10$^{-12}$ m/V, 4 × 10$^{-24}$ m$^2$/V$^2$ [25]. The Eq. (2) indicates that a large value of $E_{loc}$ in Eq. (1) facilitate the obtainment of large nonlinear excitation polarization. The intensity of the nth-order nonlinear excitation polarization $|P^{(n)}|^2$ exhibits an enhancement factor $F_R^n$ [1]. Furthermore, the total enhancement of the nth-order nonlinear response becomes $F_R^n \times \sim F_R$ when including the emission enhancement [1]. Thus, such a large value of $F_R^{n+1}$ at the HSs can aid in the realization of nonlinear spectroscopy with continuous-wave (CW) laser excitation [26,27]. Thus, it is worth to apply various HSs to CW laser excited plasmon-enhanced nonlinear spectroscopy.

In this study, the 1D HSs of NW dimers were subjected to two-photon nonlinear spectroscopy with CW near-infrared (NIR) laser excitation. We found that the two-photon-excited emissions of dye molecules, such as hyper-Rayleigh, hyper-Raman, and two-photon fluorescence, were generated solely from the edge of 1D HSs and propagated through 1D HSs. The numerical EM calculation of $F_R$ along the 1D HSs revealed that the $F_R$ of the NIR region at the edges of the 1D HSs was considerably larger than the $F_R$ at the centers by approximately $10^2$ times, resulting in these two-photon excited emissions being observed only from the edges. Analysis of the NW dimer gap distance ($d_{gap}$) dependence of $F_R$ along 1D HSs revealed that the origin of the large $F_R$ at edges in the NIR region was the lower branch of the monopole-monopole (MM) coupled SP mode that was compressed and localized at the edges for $d_{gap} = 0$ nm. The propagation of two-photon-excited emissions was supported by a higher-order coupled SP mode between the higher branch of the MM-coupled mode and the lower branch of the dipole-dipole (DD)-coupled SP mode.

**II. Theoretical and experimental background**

We explained the enhancement factors of linear and lowest-order nonlinear optical processes, which include photoemissions such as SEF, surface-enhanced Rayleigh scattering (SE-Ray), SERS, surface-enhanced two-photon fluorescence (two-photon SEF), surface-enhanced hyper-Rayleigh scattering (SEH-Ray), and surface-enhanced hyper Raman scattering (SEHRS) under the dipole approximation, and $F_R$ in Eq. (1). The availability of the dipole approximation for HSs was experimentally confirmed by the spectral similarity between the absorption and $F_R$ using silver NP dimer HSs emitting SERS signals [9].

The Hamiltonian describing a set of molecular electrons interacting a EM field is expressed as

$$\hat{H} = \hat{H}_e + \hat{H}_{ph} + \hat{H}_{int}, \quad (3)$$

where $\hat{H}_e$ and $\hat{H}_{ph}$ are the free Hamiltonians of the electrons and EM field, respectively. $\hat{H}_{int}$, which indicates the EM interaction between the electrons and the field, is described

as follows:

$$\hat{H}_{int} = \sum_{i=1}^{N} -\frac{e_i}{m_i}\hat{\mathbf{A}}(\mathbf{r}_i)\cdot\hat{\mathbf{p}}_i + \sum_{i=1}^{N} \frac{e_i^2}{2m_i}\hat{\mathbf{A}}^2(\mathbf{r}_i), \quad (4)$$

where $e_i$, $m_i$, and $\mathbf{r}_i$ are the charge, mass, and position of $i$th electron, respectively [28], $\hat{\mathbf{p}}_i$ is the momentum operator of the $i$th electron as $\hat{\mathbf{p}}_i \equiv \frac{\hbar}{\mathbf{i}}\nabla_i$, and $\hat{\mathbf{A}}(\mathbf{r}_i)$ is the vector potential operator in $\mathbf{r}_i$. The first and second terms in Eq. (4), are referred to as $\hat{H}_{int}(AP)$ and $\hat{H}_{int}(A^2)$, respectively. The time-dependent Schrödinger equation of a system that interacts with molecular electrons and an EM field is described as

$$\hat{H}\Psi(\mathbf{r},t) = \mathbf{i}\hbar \frac{\partial \Psi(\mathbf{r},t)}{\partial t}, \quad (5)$$

where $\Psi(\mathbf{r},t)$ is the wave function of the entire system. For $\hat{H}_e + \hat{H}_{ph} \gg \hat{H}_{int}$, we obtain a solution to Eq. (5) using perturbation theory. The time-independent Schrödinger equation for the stationary state of a system without the $\hat{H}_{int}$ in Eq. (3) can be expressed as follows:

$$\left(\hat{H}_e + \hat{H}_{ph}\right)\Psi_0^n(\mathbf{r}) = E_0^n \Psi_0^n(\mathbf{r}), \quad n = 1, 2, 3, \cdots \quad (5)$$

where $\Psi_0^n(\mathbf{r})$ and $E_0^n$ are the wave function and energy of the system in the $n$th state, respectively, and the subscript 0 denotes that these quantities are associated with an unperturbed system. In perturbation theory, $\Psi(\mathbf{r},t)$ is described as a series of $e^{-\mathbf{i}E_0^n t/\hbar}\Psi_0^n(\mathbf{r})$, including time-dependent $e^{-\mathbf{i}E_0^n t/\hbar}$, as

$$\Psi(\mathbf{r},t) = \sum_n b_n(t) e^{-iE_0^n t/\hbar} \Psi_0^n(\mathbf{r}), \quad (6)$$

where $b_n(t)$ is the amplitude of $\Psi(\mathbf{r},t)$. By assuming that the system initially in $\Psi_0^i$ at $t=0$ begins to be perturbed by $\hat{H}_{int}$, the $|b_f(t)|^2$ of the first- to third-order perturbation terms are derived as

$$|b_f(t)|^2 \propto \left| \langle \Psi_0^f | \hat{H}_{int} | \Psi_0^i \rangle + \sum_{n \neq i} \langle \Psi_0^f | \hat{H}_{int} | \Psi_0^n \rangle \langle \Psi_0^n | \hat{H}_{int} | \Psi_0^i \rangle \right. $$
$$\left. + \sum_{m,n \neq i} \langle \Psi_0^f | \hat{H}_{int} | \Psi_0^n \rangle \langle \Psi_0^n | \hat{H}_{int} | \Psi_0^m \rangle \langle \Psi_0^m | \hat{H}_{int} | \Psi_0^i \rangle \right|^2, \quad (7)$$

where $\langle \Psi_0^f | \hat{H}_{int} | \Psi_0^i \rangle$, $\sum_{n \neq i} \langle \Psi_0^f | \hat{H}_{int} | \Psi_0^n \rangle \langle \Psi_0^n | \hat{H}_{int} | \Psi_0^i \rangle$, and $\sum_{m,n \neq i} \langle \Psi_0^f | \hat{H}_{int} | \Psi_0^n \rangle \langle \Psi_0^n | \hat{H}_{int} | \Psi_0^m \rangle \langle \Psi_0^m | \hat{H}_{int} | \Psi_0^i \rangle$ indicate the transitions related to the first- to third-order perturbation terms, respectively, and the subscripts $m, n$ denote the intermediate states.

We now discuss the transitions induced by $\langle \Psi_0^f | \hat{H}_{int} | \Psi_0^i \rangle$ in Eq. (7). This matrix element is separated into $\langle \Psi_0^f | \hat{H}_{int}(AP) | \Psi_0^i \rangle$ and $\langle \Psi_0^f | \hat{H}_{int}(A^2) | \Psi_0^i \rangle$ using Eq. (4). Considering dipole approximation, the former term corresponding to one-photon absorption or one-photon emission is well-known as Fermi's golden rule and is described as follows:

$$\langle \Psi_0^f | \hat{H}_{int}(AP) | \Psi_0^i \rangle \propto \sum_{i=1}^N \frac{\mathbf{i}}{\hbar} (\varepsilon_f - \varepsilon_i) \langle f | e_i \mathbf{r}_i | i \rangle, \quad (8)$$

where $|i\rangle$ and $|f\rangle$ are the eigenfunctions of initial and final states of the electron system,

respectively, and $\varepsilon_i$ and $\varepsilon_f$ are their energies, respectively [9,21]. Fluorescence comprises one-photon absorption and emission transitions; thus, the two-step process of Eq. (8) is mediated by a vibrational relaxation process, as shown in Fig. 1(a1). Thus, the enhancement factor of the SEF is described as $F_R(\lambda_{ex})F_R(\lambda_{em})$, where $\lambda_{ex}$ and $\lambda_{em}$ are the excitation and emission wavelengths, respectively. Fluorescence is also quenched by energy transfer from the excited state to the metal surface during vibrational relaxation [1,29,30]. Thus, the total enhancement factor of the SEF is expressed as $F_R(\lambda_{ex})F_R(\lambda_{em})/F_Q$, where $F_Q$ is the quenching factor [1,29,30]. $\langle f|e_i\mathbf{r}_i|i\rangle \neq 0$ requires $|i\rangle \neq |f\rangle$, which indicates that Eq. (8) cannot contribute to the Rayleigh scattering. The latter term is described as follows:

$$\langle \Psi_0^f | \hat{H}_{int}(A^2) | \Psi_0^i \rangle \propto \sum_{i=1}^{N} \frac{\mathbf{i}}{\hbar} \langle f\|i\rangle. \quad (9)$$

$\langle f\|i\rangle$ can be non-zero for $|i\rangle = |f\rangle$, implying that the two-photon process in Eq. (9) corresponds to the Rayleigh scattering, as shown in Fig. 1(a2). Thus, the enhancement factor of the SE-Ray is described as $F_R^2(\lambda_{ex})$.

We discuss these transitions using the second-order perturbation term $\sum_{n \neq i} \langle \Psi_0^f | \hat{H}_{int} | \Psi_0^n \rangle \langle \Psi_0^n | \hat{H}_{int} | \Psi_0^i \rangle$ in Eq. (7). The matrix elements corresponding to Rayleigh scattering, Raman scattering, and two-photon absorption should include $\hat{H}_{int}(AP)$ as non-zero. Thus, the term can be rewritten as

$$\sum_{n \neq i} \left\langle \Psi_0^f \middle| \hat{H}_{\text{int}}(AP) \middle| \Psi_0^n \right\rangle \left\langle \Psi_0^n \middle| \hat{H}_{\text{int}}(AP) \middle| \Psi_0^i \right\rangle. \quad (10)$$

The matrix elements in Eq. (10) can contribute to Rayleigh scattering for $|i\rangle = |f\rangle$, and Raman scattering or two-photon absorption for $|i\rangle \neq |f\rangle$. A diagram of the Raman scattering is shown in Fig. 1(a3). The Rayleigh scattering intensity in Eq. (10), generated by the second perturbation term, is significantly weaker than that in Eq. (9). Thus, the matrix element in Eq. (10) primarily contributes to Raman scattering and two-photon absorption. Therefore, the enhancement factors for SERS are $F_R(\lambda_{\text{ex}})F_R(\lambda_{\text{em}})$. Two-photon fluorescence comprises two-photon absorption and one-photon emission mediated by a vibrational relaxation process. Thus, the two-step process of Eqs. (10) and (8) are considered, as shown in Fig. 1(b1). Therefore, the total EM enhancement factor for two-photon SEF is described as $F_R^2(\lambda_{\text{ex}})F_R(\lambda_{\text{em}})/F_Q$.

We discuss the transitions using the third-order perturbation term $\sum_{m,n \neq i} \left\langle \Psi_0^f \middle| \hat{H}_{\text{int}} \middle| \Psi_0^n \right\rangle \left\langle \Psi_0^n \middle| \hat{H}_{\text{int}} \middle| \Psi_0^m \right\rangle \left\langle \Psi_0^m \middle| \hat{H}_{\text{int}} \middle| \Psi_0^i \right\rangle$ in Eq. (7). The matrix element in Eq. (10) generates hyper-Rayleigh and hyper-Raman scattering. Under the dipole approximation, this matrix element should include $\hat{H}_{\text{int}}(AP)$ to be non-zero as

$$\sum_{n,m \neq i} \left\langle \Psi_0^f \middle| \hat{H}_{\text{int}}(AP) \middle| \Psi_0^n \right\rangle \left\langle \Psi_0^n \middle| \hat{H}_{\text{int}}(AP) \middle| \Psi_0^m \right\rangle \left\langle \Psi_0^m \middle| \hat{H}_{\text{int}}(AP) \middle| \Psi_0^i \right\rangle. \quad (11)$$

The matrix element in Eq. (11) can contribute to the hyper-Rayleigh scattering for $|i\rangle = |f\rangle$ and hyper-Raman scattering for $|i\rangle \neq |f\rangle$, as shown in Figs. 1(b2) and 1(b3), respectively. Hyper-Rayleigh and hyper-Raman scattering comprise three transitions: two-photon excitation and one-photon emission. Thus, the total EM enhancement factors for the SEH-Ray and SEHRS are described as $F_R^2(\lambda_{ex})F_R(\lambda_{em})$.

Figure 1(c) shows the typical photoemission spectrum of the rhodamine 6G (R6G) dye molecules enhanced by the LP resonance inside the HS of the silver NP dimer [27]. SE-Ray, SERS, and SEF spectra were observed. The SE-Ray line is considerably stronger than SERS because SE-Ray is primarily generated by the first perturbation term, whereas SERS is generated by the second perturbation term, as shown in Figs. 1(a2) and 1(a3). The SEF appearing as a broad background in Fig. 1(c) is largely quenched by $F_Q$, whose value is typically larger than that of $F_R$ by approximately $10^2$ times inside the HSs [29,30], resulting in an intensity comparable to that of SERS. Figure 1(d) shows the typical two-photon-excited emission spectra of R6G dye molecules enhanced by LP resonance inside the HS of a silver NP dimer [27]. SEH-Ray, SEHRS, and two-photon SEF spectra were observed. The SEH-Ray intensity is comparable to the SEHRS intensity because both are generated third-perturbation terms, as shown in Figs. 1(b2) and 1(b3), unlike SE-Ray and SERS. The two-photon SEF spectrum appearing as a broad background in Fig. 1(d) may also be largely quenched by $F_Q$, as shown in Fig. 1(b1), resulting in an intensity comparable to that of the SEHRS.

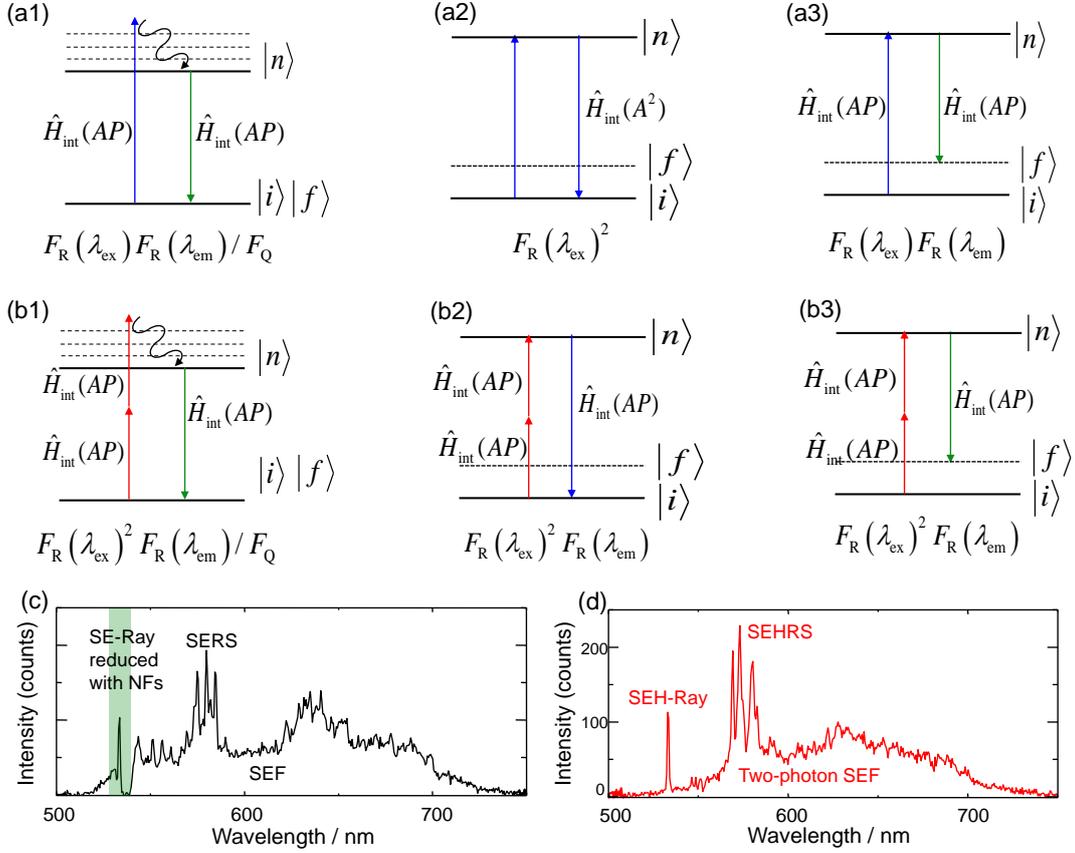

FIG. 1. Energy-level diagrams of photo-induced transitions with their total enhancement factors below the diagrams. (a1) Fluorescence process composed of absorption, vibrational relaxation (wavy arrow), and emission transitions from $|i\rangle$ to $|f\rangle$ via $|n\rangle$ triggered by two $\hat{H}_{int}(AP)$ terms. Dashed lines indicate vibrational excited states. (a2) Rayleigh scattering transition from $|i\rangle$ to $|f\rangle$ ($|i\rangle = |f\rangle$) by one $\hat{H}_{int}(A^2)$ term. (a3) Raman process composed of absorption and emission transitions from $|i\rangle$ to $|f\rangle$ via $|n\rangle$ by two $\hat{H}_{int}(AP)$ terms. Dashed line indicates vibrational excited state $|f\rangle$. (b1) Two-photon fluorescence process composed of two-photon absorption, vibrational relaxation (wavy arrow), and emission transitions from $|i\rangle$ to $|f\rangle$ via $|n\rangle$ by three $\hat{H}_{int}(AP)$ terms. Dashed lines indicate vibrational excited states. (b2) Hyper-Rayleigh scattering transition from $|i\rangle$ to $|f\rangle$ ($|i\rangle = |f\rangle$) via $|n\rangle$ composed of two-photon absorption and emission transitions by three $\hat{H}_{int}(AP)$ terms. (b3) Hyper-Raman process composed of two-photon absorption and emission transitions from $|i\rangle$ to $|f\rangle$ via $|n\rangle$ by three $\hat{H}_{int}(AP)$ terms. Dashed line indicates vibrational excited state $|f\rangle$. (c) Typical plasmon-enhanced spectrum composed of SEF, SE-Ray (reduced with two notch filters (NFs) by $>10^{-8}$ times), SERS of R6G molecules inside a HS of silver NP dimer. (d) Typical plasmon-enhanced two-photon excited spectrum composed of two-photon SEF, SEH-Ray, SEHRS of R6G molecules inside a HS of silver NP dimer.

**III. Experimental**

The silver NW colloidal dispersion was prepared as described previously [31]. The average NW diameter and length were 70 nm and 10 μm, respectively. Figures 2(a) and 2(b) show a scanning electron microscopy (SEM) image and NW diameter ($D_{NW}$) distribution, respectively. A mixture of the NW dispersion and rhodamine 6G (R6G) methanol solution (approximately $5.0 \times 10^{-6}$ M) was dropped and dried on a glass plate. Thereafter, the effective concentration of the dye on the NWs was reduced by photobleaching most of the excess dye molecules adsorbed on the NWs and glass surface using green laser beam excitation for approximately 20 min, which was confirmed by the photobleaching of fluorescence from the glass surface. The quality of the NWs was examined via Fabry–Perot interference of the SP mode [20,32], indicating that the NWs can be treated as single crystalline rather than polycrystalline. We selectively measured the two-photon emission of the NW dimers. The selection of dimers, excluding higher-order aggregates, was performed using dark- and bright-field observations. Figures 2(c) and 2(d) show dark- and bright-field images of the NW aggregates. The dimer and trimer positions can be identified using the color steps in the images.

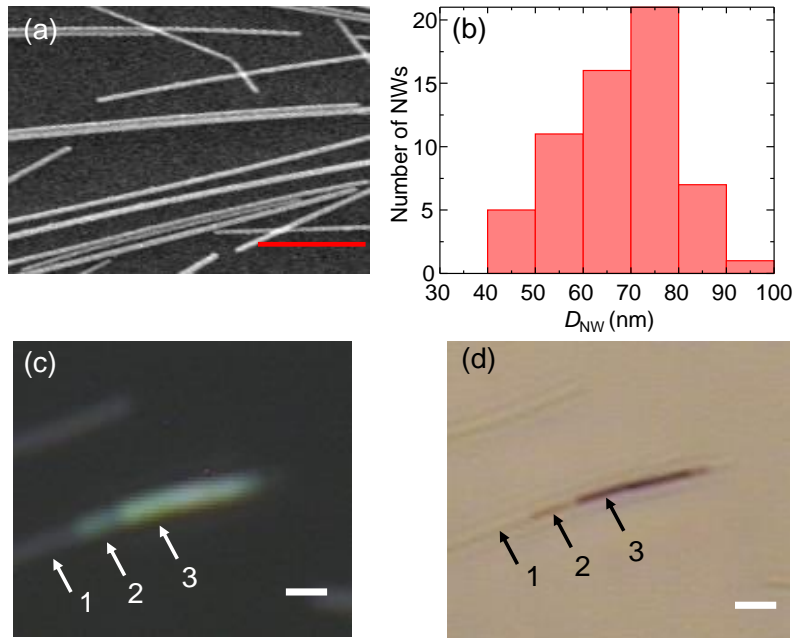

FIG. 2. (a) SEM image of NWs and NW dimers. (b) Distribution of $D_{NW}$s estimated by SEM images. (c) and (d) Dark- and bright- field images of NW trimer. The number of NWs is indicated by the arrows. All scale bars are commonly 1 μm.

The experimental setups for Rayleigh scattering, SERS (including SEF), and SEHRS (including SHE-Ray and two-photon SEF) are shown in Figs. 3(a)–3(c), respectively. Figure 3(a1) illustrates the illumination through a dark-field condenser [numerical aperture (NA) 0.92] using a 50-W halogen lamp as a white light source for measuring the Rayleigh scattering spectrum of the NW dimers. Figure 3(a2) shows a dark-field image of the NW dimers and monomers on a glass plate. The dimers can be identified by their vivid colors because the monomers always exhibit a gray color. Figure 3(b1) illustrates the wide-field illumination (approximately $200 \times 300$ μm$^2$) by a green laser beam (DPSS laser, 532 nm) for the excitation of SERS and SEF. The laser beam was focused onto the sample using a lens (NA = 0.2). The power density of the laser beam was 35 W/cm$^2$ at the focal point. Figure 3(b2) shows a SERS image of the NW dimer. The dimeric region in Fig. 3(a2) exhibits SERS activity, as shown in Fig. 3(b2). Figure 3(c1) illustrates narrow-field illumination (approximately 0.65 × 0.65 μm$^2$) by a CW-NIR beam (cw-Nd3+:YAG laser, 1064 nm) for excitation of SEHRS including SHE-Ray and two-photon SEF. The NIR laser beam was focused on the sample using an objective lens (× 100, NA 1.3). The power density of the NIR laser beam was 3.0 × 10$^6$ W/cm$^2$ at the focal point. Figure 3(c2) shows a SEHRS image of the 1D HS. The edge of the 1D HS in Fig. 3(b2) exhibits SEHRS activity, as shown in Fig. 3(c2).

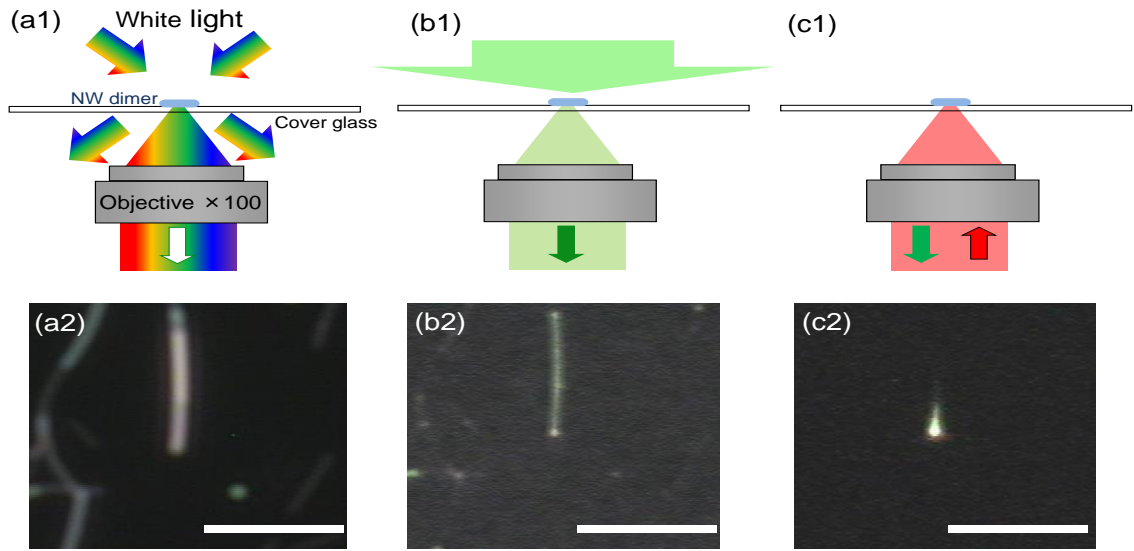

FIG. 3 Experimental setups for Rayleigh scattering, SERS including SEF, and SEHRS including SHE-Ray and two-photon SEF. (a1) Dark-field excitation with white light for detecting Rayleigh scattering light of NW dimers. (a2) Dark-field image of the NWs and NW dimers. The NWs and NW dimers placed on the cover glass are excited from above and forward Rayleigh scattering light from single NW dimer is detected with the objective. (b1) Wide-field excitation with green laser beam for detecting SERS including SEF of the NWs and NW dimers. (b2) SERS image of the NW dimer. The NW dimer is excited from above and forward SERS light is detected with the common objective. (c1) Narrow-field excitation with NIR laser beam for detecting SEHRS including SHE-Ray and two-photon SEF of single NW dimer. (c2) SEHRS image of the NW dimer. The NW dimer is excited from below and forward SEHRS light is detected with the common objective. All scale bars are 5 μm.

The Rayleigh scattering, SERS, and SEHRS light from the identical 1D HS of the NW dimer were collected through a common objective lens by adjusting its NA from 0.6-1.3 [33]. The collected light was then sent to a polychromator equipped with a thermoelectrically cooled charge-coupled device assembly. The intensity of the Rayleigh scattering was divided by the spectrum of the white light source to convert the scattering intensity into its efficiency.

**III. Results and discussion**

Figures 4(a1)–4(a3) and 4(b1)–4(b3) show the Rayleigh scattering spectra and dark-field images of the NW dimers. The Rayleigh scattering spectra exhibited a LP resonance maximum wavelength $\lambda_{LP}$ of approximately 570 nm. These results are consistent with previous reports that LP resonances with polarization directions perpendicular to the long axes of NW dimers determine the spectral features of Rayleigh scattering [20-22]. Figures 4(c1)–4(c3) show the SERS images including the SEF of the dimers. The dimer regions in Figs. 4(b1)–4(b3) exhibit SERS activity, as in Figs. 4(c1)–4(c3), indicating that the 1D HSs are generated by the $F_R$ of the LP resonances [20]. However, the edges of the 1D HSs show brighter SERS emission than their centers, indicating that other plasmon modes

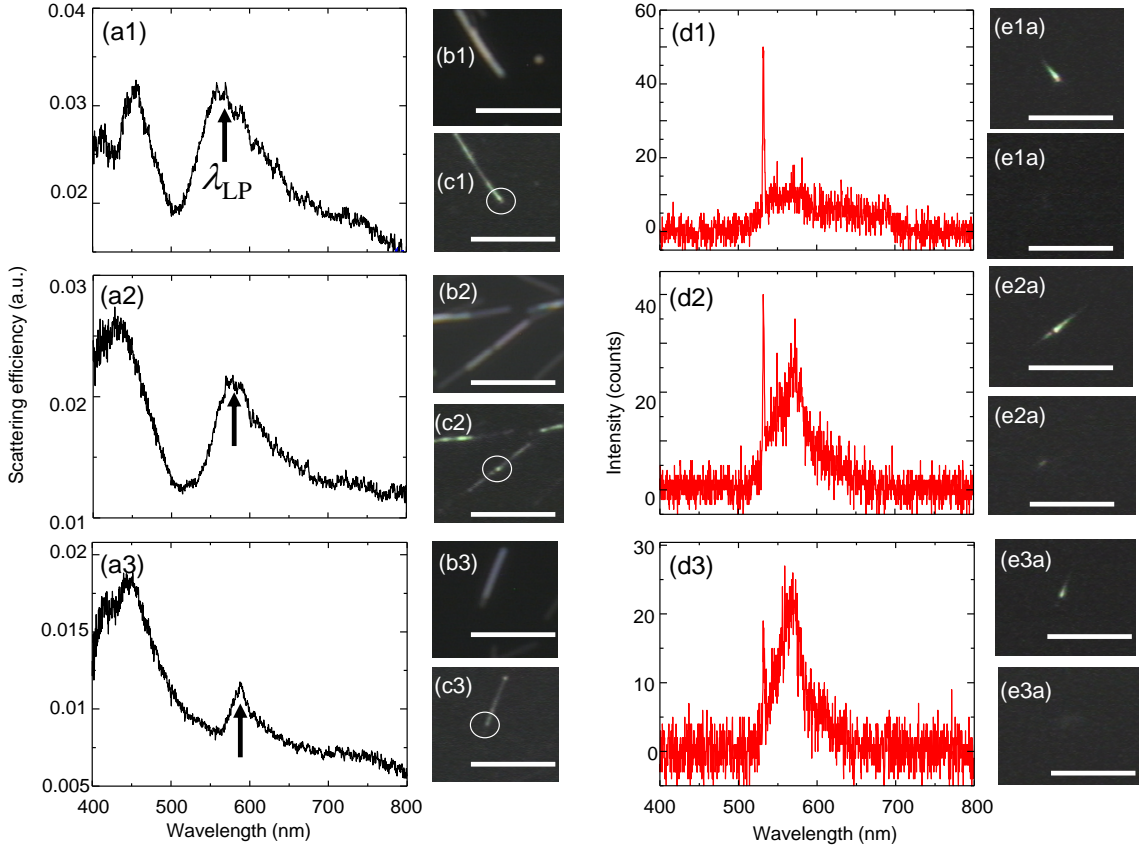

FIG. 4 (a1)–(a3) Rayleigh scattering spectra of the single NW dimers. $\lambda_{LP}$ is indicated by arrows. (b1)–(b3) Dark-field images of single isolated NW dimers. (c1)–(c3) SERS and SEF images of the single NW dimers showing 1D HSs. The edges of 1D HSs are indicated by white open circles. (d1)–(d3) Spectra of SEHRS including SHE-Ray and two-photon SEF for the single NW dimers with excitation of 1D HS edges. (e1a)–(e3a) and (e1b)–(e3b) Images of SEHRS including SHE-Ray and two-photon SEF of single NW dimers with excitation of the 1D HS edges and centers, respectively. Scale bars of all images are commonly 5 μm.

may contribute to the $F_R$ at the edges. We also reported that $F_R$ at the edges was redshifted from that around the centers, supporting this indication [22]. Figures 4(d1)–4(d3) show the variations in the two-photon emission spectra measured by exciting the edges of the 1D HSs. Figure 4(d1) shows the SHE-Ray spectrum. By contrast, Fig. 4(d3) shows the SEHRS and two-photon SEF spectra. Such spectral variations have been discussed as a result of spectral modulation by $F_R(\lambda_{em})$ in Eq. 1 [27,34]. Figures 4(e1a)–4(e3a) and 4(e1b)–4(e3b) show the two-photon emission images obtained by exciting the edges and centers of the 1D HSs, respectively. Only the edges generate strong two-photon emissions, indicating that $F_R(\lambda_{ex})$ or $F_R(\lambda_{em})$ at the edges was considerably larger than $F_R(\lambda_{ex})$ or $F_R(\lambda_{em})$ around the centers. The propagation of the two-photon emission through the 1D HSs is also shown in Figs. 4(e1a)–4(e3a).

Figure 5 shows the relationship between two-photon emission intensities and $\lambda_{LP}$ for of the edges and centers of 1D HSs for 35 NW dimers. The value of intensities were obtained from the SEHRS images such as those in Figs. 4(e1a)–4(e3a) and 4(e1b)–4(e3b). The vertical axis represents the logarithmic scale. The intensities at the edges are always larger than the intensities at the centers by $10^2$ to $10^3$ times. The gray band in Fig. 5 indicates that it is difficult to distinguish between the noise and signal owing to the disturbance of the emission of the glass plate. Thus, the two-photon emission intensities at the centers are almost negligible. The tendency for a far stronger two-photon emission at the edges indicates that the plasmon resonance generating $F_R$ at the edges is significantly different from the LP resonance at the centers that generate SERS and SEF.

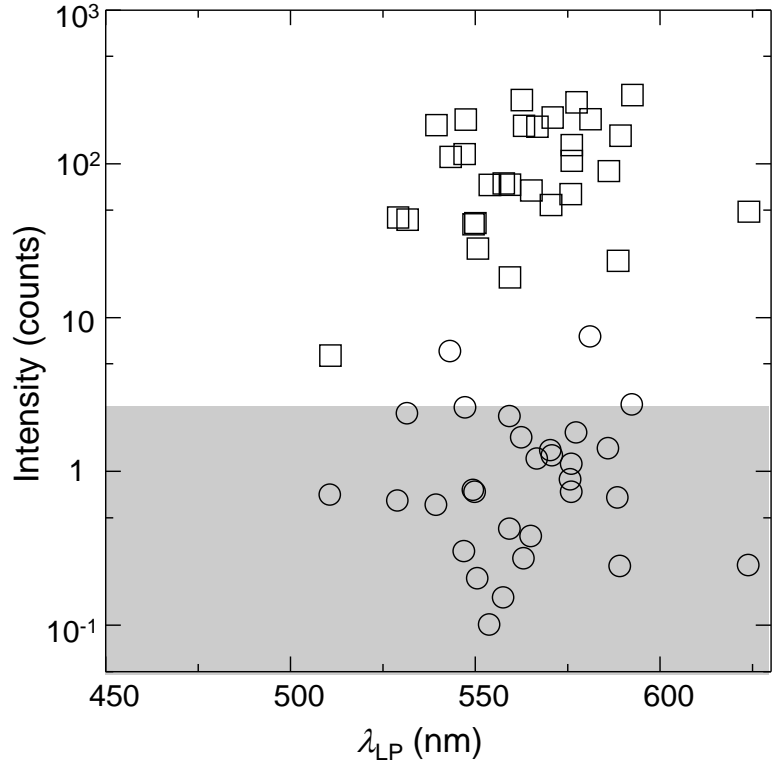

FIG. 5 Relationship between $\lambda_{LP}$s and two-photon emission intensities including SEHRS, SHE-Ray, and two-photon SEF for the edges (□) and centers (○), respectively, of 1D HSs for 35 NW dimers. The gray band indicates the area difficult for distinguishing between the noise and signal due to the disturbance of emission of the glass plate.

We performed numerical calculations for $F_R$ distributions along the 1D HSs to clarify the tendency shown in Fig. 5 using the finite-difference time-domain (FDTD) method (EEM-FDM Version 5.1, EEM Co., Ltd., Japan). The complex refractive indices of the silver NWs were adopted from a previous study [35]. The effective refractive index of the surrounding medium was set to 1.25 to ensure consistency between the calculated and experimental Rayleigh scattering spectra of the gold NPs [21]. In these calculations, the nonlocal effect, which reduces $F_R$ via Landau damping owing to unscreened surface electrons [36], was not considered because Landau damping does not change the spectral shape of $F_R$ but rather its intensity [29].

Figure 6(a) shows the setup for the FDTD calculation of the NW dimer in the coordinate system. The beam diameter of the excitation light was infinite. Thus, the contributions of both the LP and SP resonances are included in $F_R$. The excitation polarization direction was set perpendicular to the long axis of the NW dimer, because an intense $F_R$ of the 1D HS was generated along this direction by the coupled LP resonance polarization of both NWs [20-22]. The NW dimer with $D_{NW}$ comprises two cylindrical NWs with hemispherical ends. The gap distance $d_{gap}$ was set to 0 nm based on the SEM images of the dimers [20]. The lengths of the two NWs were 5.5 and 4.5 μm, respectively. Thus, the length of the 1D HS was 4.5 μm. The amplitude of the incident electric field $|E_{in}|$ is was set to be 1.0 V/m. Thus, the value of the calculated amplitude $|E_{loc}|$ was the same as that of the $F_R$ as $|E_{loc}/E_{in}| = \sqrt{F_R}$. Figure 6(b1) shows an x-z plane image of the $\sqrt{F_R}$ of the NW dimer with $D_{NW}$ = 60 nm at $\lambda_{ex}$ of 525 nm. The value of y was set to 9 nm, where $\sqrt{F_R}$ was maximum at z = 2.0 μm. The 1D HS was clearly visualized along the junction of the NW dimer. Figure 6(b2) shows the profile of $\sqrt{F_R}$ along the z-axis at x, y = 0 and 9 nm at $\lambda_{ex}$ of 525 nm. The oscillating structure is the result of the Fabry–Perot interference of a SP wave [20]. Figure 6(c) shows the $\sqrt{F_R(\lambda_{ex})}$ spectra along a

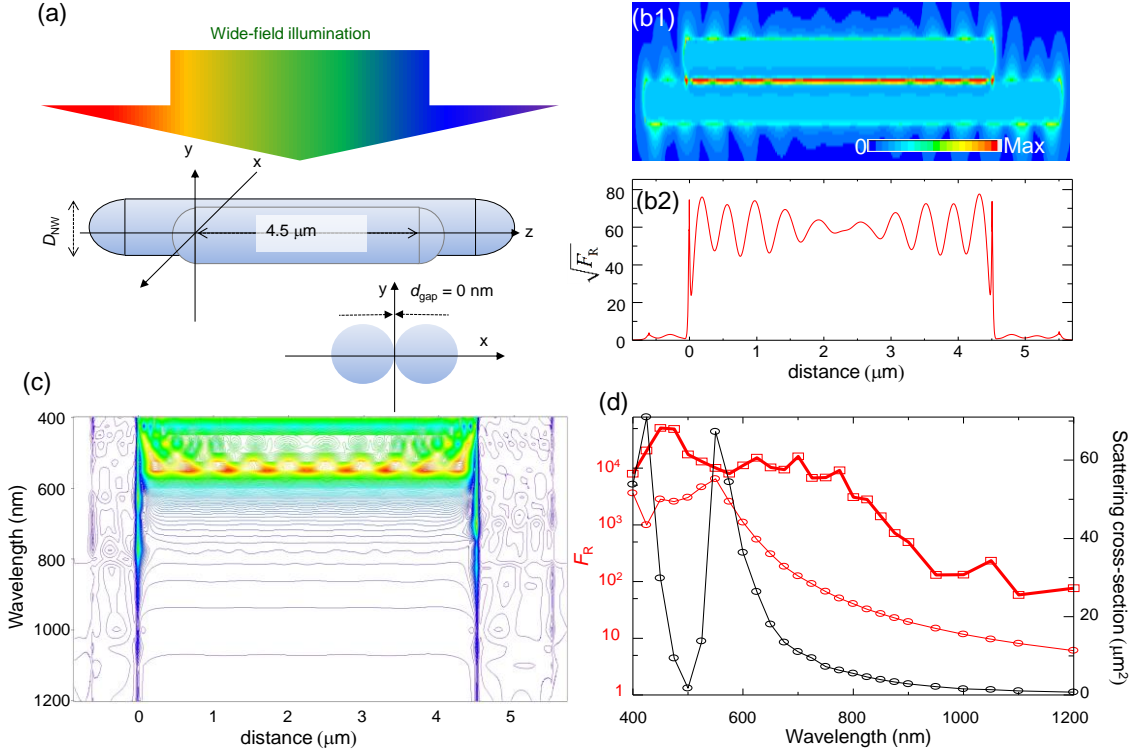

FIG. 6 (a) FDTD calculation setup for $|E_{loc}/E_{in}|(=\sqrt{F_R})$ around single NW dimer by wide-field white light excitation from upper side with the coordinate system. Polarization direction of excitation light are orthogonal to the direction of 1D HS. (b1) $\sqrt{F_R}$ distributions along a 1D HS of x-z plane at y = 9 nm at $\lambda_{ex}$ = 525 nm. (b2) $\sqrt{F_R}$ profile along z-axis at x, y = 0, 9 nm at $\lambda_{ex}$ of 525 nm. (c) $\sqrt{F_R(\lambda_{ex})}$ spectra along a 1D HS from $\lambda_{ex}$ of 400 to 1200 nm expressed as a contour map. (d) Rayleigh scattering spectrum of the NW dimer (black curve with ○) and the $F_R(\lambda_{ex})$ spectra at the edge (red bold curve with ○) and around the center (red thin curve with ○) of 1D HS.

1D HS from $\lambda_{ex}$ of 400–1200 nm expressed as a contour map. The $\lambda_{LP}$ of the LP resonance appeared at approximately 550 nm around the center of the 1D HS of the $\sqrt{F_R(\lambda_{ex})}$ spectra. However, the $\sqrt{F_R(\lambda_{ex})}$ spectra at the edge were largely different from those around the center and broadened toward the longer-wavelength side. Thus, this difference may be related to the experimental result that the edges solely generate two-photon excited emissions, as shown in Figs. 4 and 5. Figure 6(d) shows the Rayleigh scattering spectrum and the $F_R(\lambda_{ex})$ spectra at the edge and around the center. The values of $F_R$ around the centers were obtained by averaging the $F_R$ within z = 1–3 μm along the 1D HS. The vertical axis of the $F_R(\lambda_{ex})$ spectra is on a logarithmic scale. The spectral shape of the $F_R$ at the center of the 1D HS was similar to that of the Rayleigh scattering spectrum, indicating that the $F_R$ around the center was generated by the LP resonance of the NW dimer [20]. However, the spectral shape of the $F_R$ at the edge was different from that of the Rayleigh scattering spectrum, indicating that the $F_R$ at the edge was not generated by the LP resonance. It is clear that the $F_R$ of the edge is considerably larger than that of the center by approximately $10^2$ times at $\lambda_{ex}$ of approximately 1050 nm. This supported the experimental results in Figs. 4 and 5 that the two-photon-excited emission was solely observed from the edges.

We attempted to reproduce the experimental results in Fig. 5 using the FDTD calculations. Figures 7(a1)–7(a5) show the Rayleigh scattering spectra of NW dimer and $F_R(\lambda_{ex})$ spectra at the edges and around centers of 1D HSs by changing $D_{NW}$ from 30 to 100 nm. With increasing $D_{NW}$, $\lambda_{LPS}$ in Rayleigh scattering spectra exhibited redshifts within approximately 450–650 nm and the maxima of $F_R$ around the centers also exhibited redshifts for 450–550 nm until $D_{NW}$ of 80 nm and were maintained around the positions. This behavior of the maxima in $F_R$ was clarified as the changing of LP

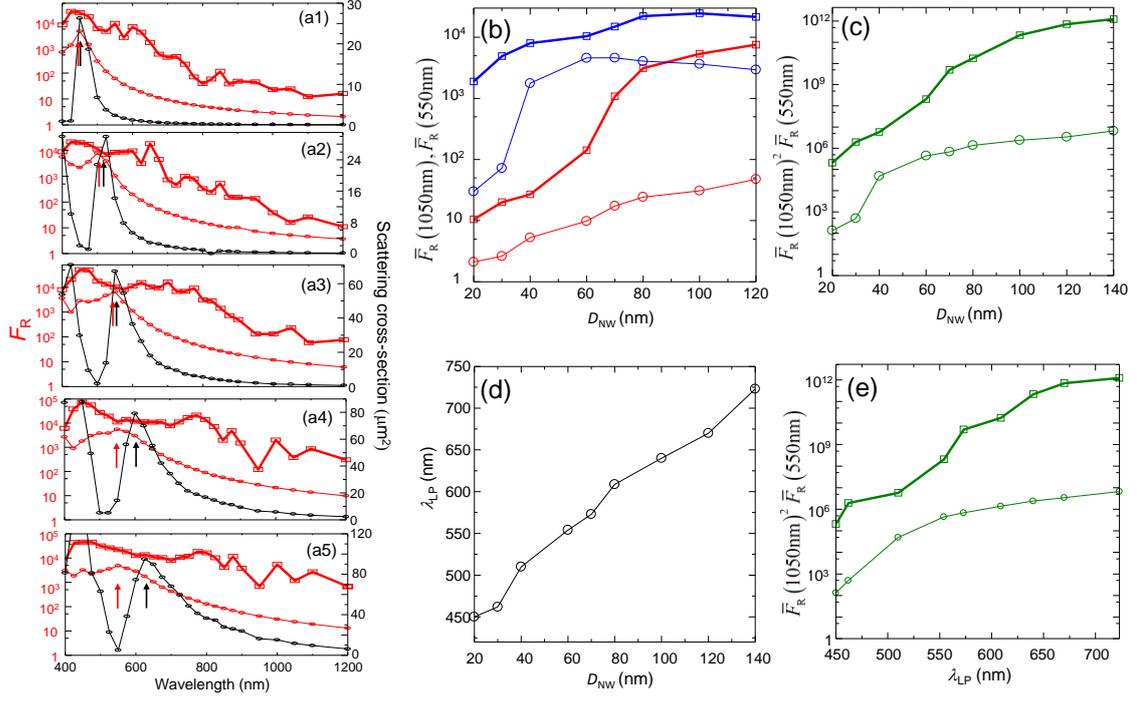

FIG. 7 (a1)–(a5) Rayleigh scattering spectra of the NW dimer (black curves with ○) and $F_R(\lambda_{ex})$ spectra at the edge (red bold curves with □) and around the center (red thin curves with ○) of 1D HSs for $D_{NW}$ = 30, 40, 60, 80, and 100 nm, respectively. Their spectral peak positions are indicated by black and red arrows, respectively. (b) $D_{NW}$ dependence of $\bar{F}_R(1050\text{nm})$ (red curves) and $\bar{F}_R(550\text{nm})$ (blue curves) at the edge (red or blue curve with □) and around the center (red or blue curve with ○) of 1D HSs. (c) $D_{NW}$ dependence of $\bar{F}_R(1050\text{nm})^2 \bar{F}_R(550\text{nm})$ at the edge (green bold curve with □) and around the center (green thin curve with ○) of 1D HSs. (d) $D_{NW}$ dependence of $\lambda_{LP}$ of NW dimers. (d) $\lambda_{LP}$ dependence of $\bar{F}_R(1050\text{nm})^2 \bar{F}_R(550\text{nm})$ at the edge (green bold curve with □) and around the center (green thin curve with ○) of 1D HSs.

resonance generating $F_R$ from dipole-dipole (DD)-coupled LP mode to dipole-quadrupole (DQ)-coupled LP mode [21,37,38]. The maxima of $F_R$ at the edges exhibited spectral broadening with redshifts. This spectral behavior is significantly different from the $F_R$ around the centers, indicating that unknown plasmon mode contributes to the $F_R$ at the edges. As discussed in Sec. II, the enhancement factor of two-photon excited emission is expressed as $F_R(\lambda_{ex})^2 F_R(\lambda_{em})$. Thus, the relationship between $D_{NW}$ and $F_R$ is examined for the edges and centers of 1D HSs using $\lambda_{ex}$ and $\lambda_{em}$ of approximately 1050 and 550 nm, respectively. Figure 7(b) shows the relationship by changing $D_{NW}$ from 20 to 120 nm. Both $\bar{F}_R(1050\text{nm})$ and $\bar{F}_R(550\text{nm})$ at the edges are larger than those around the centers, where $\bar{F}_R(1050\text{nm})$ and $\bar{F}_R(550\text{nm})$ mean the averaged values of $F_R$s for $\lambda_{ex}$ and $\lambda_{em}$ of 1000–1100 and 525–575 nm, respectively. Consequently, the relationship between $D_{NW}$ and $\bar{F}_R(1050\text{nm})^2 \bar{F}_R(550\text{nm})$ was calculated using Fig. 7(b). Figure 7(c) shows the relationship by changing $D_{NW}$ from 20 to 140 nm. The values of $\bar{F}_R(1050\text{nm})^2 \bar{F}_R(550\text{nm})$ at the edges were considerably larger than those around the centers by $10^2$–$10^6$ times. Regarding the experimental distribution of $D_{NW}$ for 50–100 nm, as in Fig. 2(b), this calculation result was consistent with the experimental results in Fig. 5, demonstrating the relationship between $\lambda_{LP}$ and two-photon excited emission intensities. To confirm this consistency, the relationship between $D_{NW}$ and $\bar{F}_R(1050\text{nm})^2 \bar{F}_R(550\text{nm})$ in Fig. 7(c) is converted to that between $\lambda_{LP}$ and $\bar{F}_R(1050\text{nm})^2 \bar{F}_R(550\text{nm})$ using Fig. 7(d), which indicates the relationship between $D_{NW}$ and $\lambda_{LP}$. Figure 7(e) exhibits the relationship between $\lambda_{LP}$ and $\bar{F}_R(1050\text{nm})^2 \bar{F}_R(550\text{nm})$. This calculated relationship is consistent with experimental one in Fig. 5, thus confirming that the edges exhibit considerably stronger enhancement factors than the centers.

Figure 8 shows the calculated relationship between $\lambda_{LP}$ and $\bar{F}_R(1050\text{nm})^2 \bar{F}_R(550\text{nm})$ and the experimental relationship between $\lambda_{LP}$ and two-photon-excited emission intensities at the edges and around the centers of the 1D HSs. Both the vertical axes are linear. In the calculated relationship, $\bar{F}_R(1050\text{nm})^2 \bar{F}_R(550\text{nm})$ sharply increases from $\lambda_{LP}$ ~550 nm, which corresponds to a $D_{NW}$ of 60 nm. The experimental relationship appears to reproduce the increase from $\lambda_{LP}$ ~550 nm. The calculated and experimental relationships show that both the $\bar{F}_R(1050\text{nm})^2 \bar{F}_R(550\text{nm})$ and two-photon-excited emission intensities are negligible around the centers. The consistencies between the calculations and experiments indicate that electromagnetism can explain why the $F_R$ of the edges is considerably larger than that of the centers.

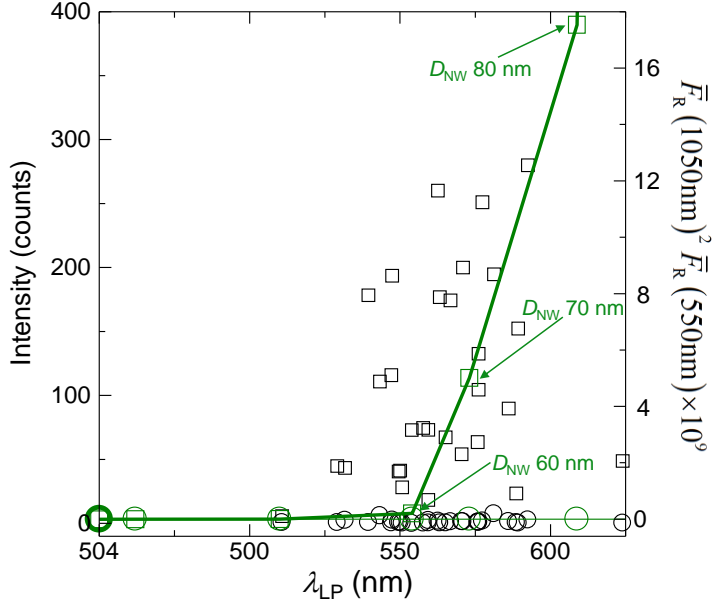

FIG. 8 Calculated relationship between $\lambda_{LP}$s and $\bar{F}_R(1050\text{nm})^2 \bar{F}_R(550\text{nm})$s for the edges (green bold curve with □) and centers (green thin curve with ○), respectively, of 1D HSs for NW dimers with $D_{NW}$ of 20 to 80 nm and experimental relationship between $\lambda_{LP}$s and two-photon emission intensities including SEHRS, SHE-Ray, and two-photon SEF for the edges (□) and centers (○), respectively, of 1D HSs of 35 NW dimers.

We explored the mechanism of strong $F_R(\lambda_{ex})^2 F_R(\lambda_{em})$ for two-photon-excited emissions at the edges of 1D HSs. Figures 7(a1)–7(a5) show that the $F_R(\lambda_{ex})$ spectra at the edges are significantly different from those around the center for $d_{gap}$ = 0 nm. To clarify this difference, we examined the $F_R(\lambda_{ex})$ spectra along the 1D HSs by changing $d_{gap}$. Figures 9(a1)–9(a5) show the contour maps of the $F_R(\lambda_{ex})$ spectra along the 1D HSs obtained by changing $d_{gap}$ from 20 to 0 nm. The oscillation patterns became finer and finally localized around the edges as decreasing $d_{gap}$. These oscillation patterns are Fabry–Perot interferences of SP waves, reflecting their wavenumbers [20]. Thus, the confinement of fields by decreasing $d_{gap}$ results in an increase in the wavenumber. SPs cannot be directly coupled to light propagating in free space because their momenta are always larger than those of light [19,22]. Thus, SPs are indirectly coupled with light via. LP at the edge [22]. Thus, the propagation of SP waves always begins at the edges. The propagation distance decreases with increasing wavenumber because the ohmic loss per distance increases. Thus, the light energy stored in the SP mode is compressed and finally localized at the edge for $d_{gap}$ = 0 nm, as shown in Fig. 9(a5). Figures 9(b1)–9(b5) and 9(c1)–9(c5) show the profiles of $\sqrt{F_R(550\text{nm})}$ and $\sqrt{F_R(1050\text{nm})}$ along the 1D HSs

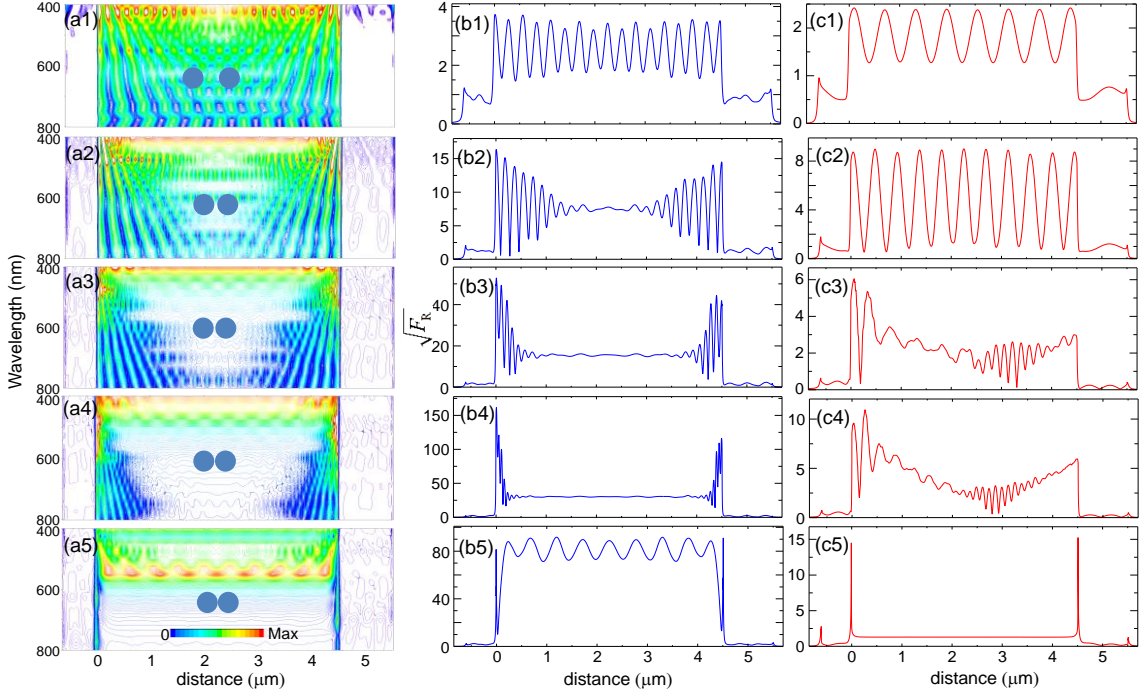

FIG. 9 (a1)–(a5) $d_g$ dependence of $\sqrt{F_R(\lambda_{ex})}$ profiles along 1D HS of the NW dimers for $d_g$ = 20, 5, 2, 1, and 0 nm, respectively, with $D_{NW}$ = 60 nm from $\lambda_{ex}$ of 400 to 800 nm expressed as contour maps. The cross-section images of NW dimers are indicated in each panel. (b1)–(b5) $\sqrt{F_R(550\text{nm})}$ profiles along a 1D HSs of the NW dimers for $d_g$ = 20, 5, 2, 1, and 0 nm, respectively, with $D_{NW}$ = 60 nm. (c1)–(c5) $\sqrt{F_R(1050\text{nm})}$ profiles along 1D HSs of the NW dimers for $d_g$ = 20, 5, 2, 1, and 0 nm, respectively, with $D_{NW}$ = 60 nm.

by changing $d_{gap}$ from 20 to 0 nm, respectively. The compression and localization of the SPs at the edge are clearly observed as decreasing $d_{gap}$. This localization effect at $d_{gap}$ = 0 nm was more pronounced for $\sqrt{F_R(1050\text{nm})}$ in Fig. 9(c5) than for $\sqrt{F_R(550\text{nm})}$ in Fig. 9(b5). This difference is owing to the to the $\sqrt{F_R(1050\text{nm})}$ at the edge being solely generated by the compressed SP, whereas $\sqrt{F_R(550\text{nm})}$ is generated by both SP and LP, whose resonance appears at approximately 550 nm in Fig. 9(a5). Therefore, the large value of $F_R$ around the NIR region at the edges, induced by compressed and localized SP waves, is the origin of the selective observation of two-photon excited emission at the edges.

Here, we attempt to identify the SP mode that induces strong $F_R(\lambda_{ex})^2 F_R(\lambda_{em})$ at the edges of 1D HSs. Theoretical studies on the $d_{gap}$ dependence of the dispersion relationships of the SP modes of the NW dimers have been reported by Myroshnychenko et al. [39] and Sun et al. [40]. Based on these reports, we investigated the SP modes generating $F_R$ at the edges and centers of 1D HSs. Figures 10(a1)–10(a5), obtained from Ref. 39 shows the $d_{gap}$ dependence of the dispersion curves. At $d_{gap}$ = 20 nm. As shown in Fig. 10(a1), the dispersion curves of the lower and upper branches of the MM coupled SP modes and the lower branches of the DD-coupled SP mode are indicated as I, II, and III, respectively [39,40]. The charge distributions for I, II, and III are shown in Fig. 10(b) [40]. With decreasing $d_{gap}$, the curves I and III shift to larger wavenumbers, as shown in Figs. 10(a2)–10(a5), owing to the compression of the SP waves inside the gaps. In contrast, dispersion curve II does not exhibit such shifts because of the lack of charge distribution inside the gap, as in Fig. 10(b). Furthermore, curves I and III, both of which are the lower branches of the coupled SP modes, shift to the lower-energy sides with decreasing $d_{gap}$ owing to the increase in coupling energy. Thus, curves II and III exhibit anti-crossing behavior, as in Fig. 10(a2)–10(a5). These changes in the dispersion curves

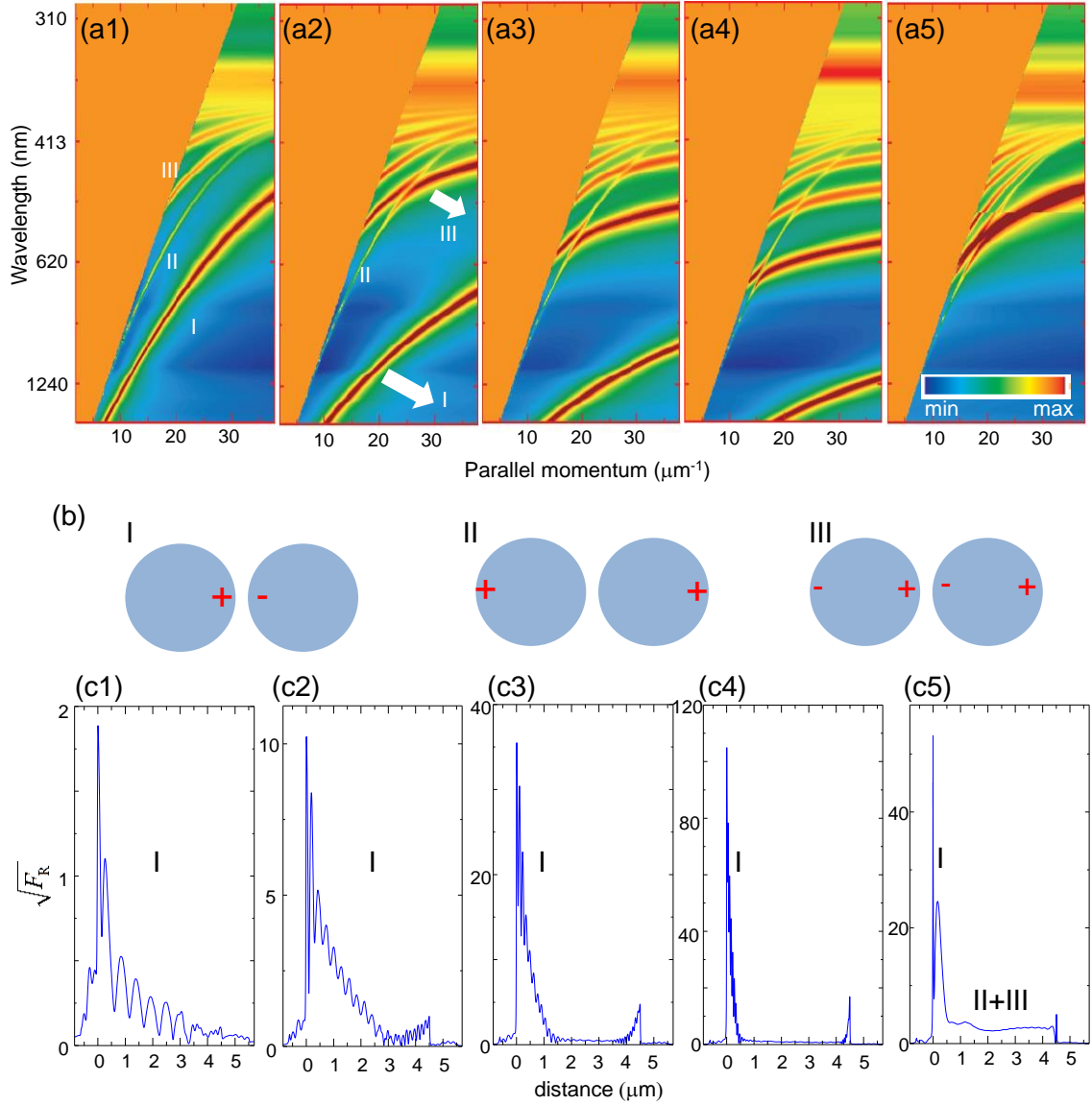

FIG. 10 (a1)–(a5) $d_g$ dependence of dispersion curves of coupled SP modes of NW dimers for $d_g$ = 20, 5, 2, 1, and -1 nm, respectively, with $D_{NW}$ = 200 nm taken from Ref. 39. In (a1), the dispersion curves of the lower and upper branches of MM coupled SP modes and the lower branches of DD-coupled SP mode are indicated by I, II, and III, respectively. (b) The charge distributions of coupled SP modes for I, II, and III in (a1) with referring Ref. 40. (c1)–(c5) $\sqrt{F_R(550nm)}$ profiles along 1D HSs of the NW dimers for $d_g$ = 20, 5, 2, 1, and -1 nm, respectively, with $D_{NW}$ = 60 nm. Note that (c1)–(c5) were calculated by excitation of the edges of 1D HSs with a light beams with diameter of 500 nm. Reprinted with permission from Ref. 39: Copyright 2012 Optical Society of America.

result in the compression and localization of the fields of the lower branches of the MM-

coupled SP mode at the edge of the 1D HS, as shown in Fig. 10(c1)–10(c4), as well as the emergence of a higher-order SP coupled mode between the upper branch of the MM-coupled modes and the lower branches of the DD-coupled mode, as shown in Fig. 10(c5). Note that Figs. 10(c1)–10(c5) were calculated by the excitation of the edges of the 1D HSs using a light beam with a diameter of 500 nm to clarify the coupled SP mode supporting the propagation of two-photon excited emissions. The lower energy shifts of curve I in Fig. 10(a2)–10(a5) resulted in a larger $F_R$ in the NIR region than in the visible region, as shown in Figs. 7(a1)–7(a5). These behaviors of the dispersion curves of the coupled SP modes reveal that the compressed MM-coupled SP mode contributes to a strong $F_R$ at the edges in the NIR region, and the propagation of the two-photon excited emission is supported by the higher-order coupled SP mode.

Finally, we quantitatively examine the FDTD calculation results and the experimental results. The hyper-Raman cross-section of a R6G molecule was estimated to be $7.8 \times 10^{-59}$ cm$^4$s/photons [26]. Thus, the power density of the NIR laser beam $3.0 \times 10^6$ W/cm$^2$ (~$1.61 \times 10^{25}$ photons/cm$^2$s) and $F_R^2(\lambda_{ex})F_R(\lambda_{em}) < 10^{12}$ in Fig. 7 generate the effective cross-section of SEHRS $< 1.26 \times 10^{-21}$ cm$^2$. This value is too small to measure SEHRS of single R6G molecule, regarding a fluorescence cross-section of a R6G molecule of approximately $10^{-16}$ cm$^2$ [13]. Thus, the large amount of R6G molecules may be on the edge of 1D HS to generate SEHRS light. Furthermore, we consider that the additional EM enhancement by fine surface roughness around the edge may contribute to boost up $F_R^2(\lambda_{ex})F_R(\lambda_{em})$. The large scatter in data points in Fig. 8 suggests this contribution.

IV. CONCLUSIONS

In this study, we investigated the SE-Ray, SEHRS, and two-photon SEF of dye molecules generated solely at the edges of 1D HSs between NW dimers with CW NIR laser excitation and propagated through 1D HSs. The FDTD calculation of $F_R$ along the 1D HSs revealed that the $F_R$ of the NIR region at the edges of the 1D HSs was considerably larger than that at the centers by approximately $10^2$ times, resulting in these two-photon excited emissions being observed only from the edges. The analysis of the $d_{gap}$ dependence of $F_R$ along 1D HSs revealed that the lower branch of the MM-coupled SP mode was compressed and localized at the edges, as the decreasing $d_{gap}$ was the origin of the large $F_R$ in the NIR region. The propagation of two-photon-excited emissions was supported by a higher-order coupled SP mode between the higher branch of the MM-coupled mode and the lower branch of the DD-coupled mode. We believe that CW laser-excited nonlinear spectroscopy using the EM enhancement of 1D HSs is applicable to various research fields such as the nonlinear counterparts of SERS, tip-enhanced Raman spectroscopy, and semiconductor-enhanced Raman spectroscopy [41-43].

ACKNOWLEDGMENTS

The authors thank to Prof. Jeyadevan Balachandran (University of Shiga Prefecture) for providing the silver nanowires.